\documentclass[webpdf,contemporary,large,namedate]{oup-authoring-template}%

\usepackage{algorithmicx}
\usepackage{lmodern}
\usepackage{setspace}

\begin{document}
\journaltitle{TBD}
\DOI{TBD}
\copyrightyear{2025}
\pubyear{2025}
\access{Advance Access Publication Date: Day Month Year}
\appnotes{Preprint}
\firstpage{1}

\title[Easy regions for short-read variant calling]{Finding easy regions for short-read variant calling from pangenome data}
\author[1,2,3,$\ast$]{Heng Li\ORCID{0000-0003-4874-2874}}
\address[1]{Department of Biomedical Informatics, Harvard Medical School, 10 Shattuck St, Boston, MA 02215, USA}
\address[2]{Department of Data Science, Dana-Farber Cancer Institute, 450 Brookline Ave, Boston, MA 02215, USA}
\address[3]{Broad Insitute of MIT and Harvard, 415 Main St, Cambridge, MA 02142, USA}
\corresp[$\ast$]{Corresponding author. \href{mailto:hli@ds.dfci.harvard.edu}{hli@ds.dfci.harvard.edu}}


\abstract{
\sffamily\footnotesize
\textbf{Background:}
While benchmarks on short-read variant calling suggest low error rate below 0.5\%,
they are only applicable to predefined confident regions.
For a human sample without such regions, the error rate could be 10 times higher.
Although multiple sets of easy regions have been identified to alleviate the issue,
they fail to consider non-reference samples or are biased towards existing short-read data or aligners.
\vspace{0.5em}\\
\textbf{Results:}
Here, using hundreds of high-quality human assemblies,
we derived a set of sample-agnostic easy regions where short-read variant calling reaches high accuracy.
These regions cover 88.2\% of GRCh38, 92.2\% of coding regions and 96.3\% of ClinVar pathogenic variants.
They achieve a good balance between coverage and easiness and can be generated for other human assemblies or species with multiple well assembled genomes.
\vspace{0.5em}\\
\textbf{Conclusion:}
This resource provides a convient and powerful way to filter spurious variant calls for clinical or research human samples.
\vspace{0.5em}\\
\textbf{Keywords:}
variant calling; pangenome; assembly
}

\maketitle

\section{Introduction}

Variant calling from whole-genome sequencing data plays a crucial role in medical genetics,
population genetics, and clinical genetic testing.
Many algorithms have been developed for accurate variant calling from short-read data~\citep{Depristo:2011vn,Kim:2018aa,Poplin:2018ab,Cooke:2021ab}
and multiple benchmarks have been conducted to evaluate their accuracy~\citep{Zook:2014ab,Li:2018aa}.

A common practice in variant calling evaluation is to compare called variants to curated ground truth,
notably generated by the Genome-In-A-Bottle (GIAB) group~\citep{Zook:2014ab}.
Benchmarks using GIAB samples often report a low error rate of $<$0.5\%~\citep{Olson:2022aa}.
On the other hand, when we compare variant calls from two callers, such as DeepVariant and GATK,
their difference often exceeds 5\%, 10 times higher than the error rate on GIAB samples.
We see this apparent discrepancy because with GIAB, we only evaluate calls in sample-specific \emph{confident regions}
where the ground truth can be trusted.
The error rate outside confident regions is much higher.
The problem is that we do not know confident regions for a clinical sample.
If we blindly take calls across the whole genome,
the error rate will be closer to 5\% than to 0.5\%.
Arguably, confident regions are more important than variant calling algorithms on GIAB-based benchmarks.

Whereas confident regions are used for variant calling benchmarks on a specific sample,
\emph{easy regions} are derived for variant filtration agnostic to samples.
They are genomic regions where general variant callers can reach high accuracy without sophisticated filtering.
At same time, we also prefer to have large easy regions such that most clinically relevant variants can be retained.
Constructing high-quality easy regions demands the balance between genome coverage and ease of calling.

Broadly speaking, there are two classes of methods to identify easy regions.
The first only considers unique regions in the reference genome.
Methods in this class differ in how to define uniqueness.
We derived a set of easy regions for ancient DNA variant filtering by selecting 35-mers that could not be mapped elsewhere within one mismatch or gap.
ENCODE constructed similar $k$-mer-based sets with GEM mapper~\citep{Marco-Sola:2012kx}
and switched to Umap~\citep{Karimzadeh:2018aa} later.
GIAB combined short-range uniqueness of GenMap~\citep{Pockrandt:2020aa}
and long-range segmental duplications~\citep{Dwarshuis:2024aa}.
It optionally excluded short tandem repeats, regions of excessively low or high GC, and X-transposed and amplicon regions on X and Y chromosomes.
Using the reference genome only, methods in this class are fast to run and easy to deploy.
However, they would overlook regions unique in the reference genome but duplicated in non-reference samples.

The second class of methods solve this problem by taking non-reference samples into account.
We inspect the read alignment of multiple samples and exclude regions of excessively high or low coverage or enriched with alignments of low mapping quality.
The 1000 Genomes Project (1KG) was the first to use such an approach~\citep{1000-Genomes-Project-Consortium:2010qc}.
Illumina posted an article in 2021 with a similar method but not citing 1KG.
\citet{Aganezov:2022aa} reproduced these regions for T2T-CHM13~\citep{Nurk:2022up}.
\citet{Pan:2022aa} went a step further by using multiple aligners and variant callers.
ENCODE blacklist~\citep{Amemiya:2019aa} considered empirical read alignment as well.
\citet{Mandelker:2016aa} defined problematic regions for GRCh37 exome which were often used in clinical testing.
Methods in this class implicitly model biases in input reads and alignment algorithms.
This helps variant calling given reads and alignment algorithms of similar property
but is biased against new data types and alternative aligners.
These methods also demand more computing and are harder to reproduce.

In this article, we will describe a new method to produce easy regions from multiple high-quality human assemblies~\citep{Liao:2023aa}.
It considers non-reference samples like the second class of methods but uses principles in the first class.
It is also not biased towards sequencing data types or existing alignment algorithms.

\section{Methods}

Suppose there are $N$ genomes including the reference genome.
A $k$-mer is \emph{sufficiently unique}
if 1) it has only one hit of edit distance up to $d_1$ in the reference genome
and 2) it has $<c\cdot N$ hits of edit distance up to $d_2$ across all genomes, $d_2>d_1$.
$c>1$ corresponds to the tolerance with copy-number gains.
With the first condition, the $k$-mer is unlikely to be mismapped under sequencing errors or mutations if the second best hit $d_1$-edit away.
The second condition kicks in when the $k$-mer is unique in the reference genome but has extra paralogous copies in non-reference samples.
In this case, the paralogous copies would be wrongly mapped to the unique reference position and
differences between paralogous copies would be called as false variants.
The second condition disregards the case where a $k$-mer is deleted in some genomes but duplicated in others.
This is a weakness of our method.

We could test the sufficient uniqueness of a reference $k$-mer by aligning it each genome separately.
This would be computationally intensive.
Instead, we construct a ropebwt3 index for all genomes including the reference~\citep{Li:2024ac}
and align a $k$-mer with the revised BWA-SW algorithm in the end-to-end mode.
This algorithm guarantees to find all exact matches and also reports suboptimal hits,
though it may miss inexact hits due to heuristics.
We sample one $k$-mer per $w$ bases from the reference genome.
If a $k$-mer at position $i$ is not sufficiently unique, we mask interval $[i,i+k)$.
After processing all sampled $k$-mers, the remaining regions constitute the easy regions.

On our dataset, $N=472$.
We use $k=151$ as 151bp reads are common nowadays.
We set $w=10$, $c=1.01$, $d_1=3$ and $d_2=7$.
Increasing $c$ to 1.02 adds 0.2\% of regions; increasing $d_2$ to 8 reduces regions by 0.1\%.
During the development of panmask, we additionally generated regions for $N=100$ and $N=398$ under a few other settings.
We found reasonable parameters generally resulted in similar regions with slightly different balance points between region size and easiness.
We also expect our current setting applicable to other species.

To save missed hits due to $k$-mer sampling and alignment heuristics,
we additionally run GenMap~\citep{Pockrandt:2020aa}, an exact algorithm though not supporting gapped alignment, with $k=151$ and maximal hamming distance 3 on the reference genome.
We drop a $k$-mer region if GenMap finds multiple hits.
We also filter out an easy region shorter than 50bp to reduce fragmentation.

We call easy regions produced above as {\sf pm151:lenient}, where ``pm'' stands for panmask.
As SNP and indel calling errors are enriched in low-complexity regions (LCRs; \citealp{Li:2014aa}),
we further exclude such regions longer than 18bp with the SDUST algorithm~\citep{Morgulis:2006aa}.
This results in {\sf pm151:strict} easy regions which are a strict subset of {\sf pm151:lenient}.

\section{Results}

We collected the following easy regions:
updated {\sf 1KG}~\citep{1000-Genomes-Project-Consortium:2010qc},
{\sf Umap100} for 100bp single-end reads~\citep{Karimzadeh:2018aa},
{\sf ENCODE} blacklist v2~\citep{Amemiya:2019aa},
reproduced Illumina regions (labeled as {\sf Ilmn-easy}; \citealp{Aganezov:2022aa}),
highly reproducible regions (labeled as {\sf highRepro}; \citealp{Pan:2022aa})
and {\sf GIAB} genome stratification~\citep{Dwarshuis:2024aa}.
Both {\sf 1KG} and {\sf GIAB} provide relaxed and strict versions.
To the best of our knowledge, these are all the available easy regions on GRCh38.
We additionally included confident regions {\sf HG002-old} from GIAB ground truth v4.2.1
and {\sf HG002-Q100} from GIAB's new Q100 benchmark v1.1.

\begin{figure}[tb]
\centering
\includegraphics[width=0.9\columnwidth]{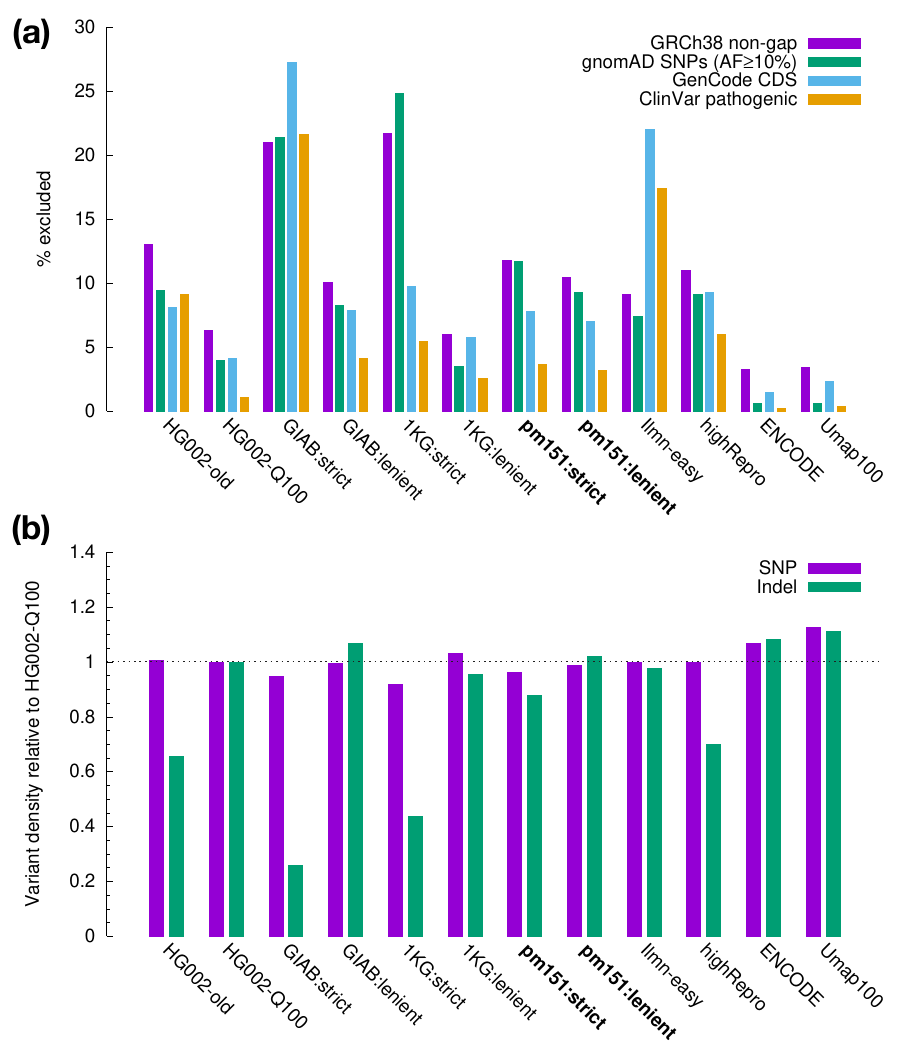}
\caption{Properties of easy and confident regions.
{\bf (a)} Coverage on GRCh38 chromosomal sequences.
The gnomAD v4.1 coverage is calculated from SNPs of allele frequency $\ge$10\% intersected with HPRC SNPs.
CDS coverage is measured with MANE-selected protein coding genes from GenCode v48.
ClinVar coverage is measured on pathogenic and likely pathogenic variants from ClinVar 20250623.
{\bf (b)} Variant density relative to HG002-Q100.
The variant density of an easy region set is the number of HG002-Q100 variant calls in the easy regions divided by the total region length.
The relative density equals to the density of the set over the density in HG002-Q100 confident regions.
}\label{fig:cov}
\end{figure}

We calculated the coverage of each set of regions (Fig.~\ref{fig:cov}a).
ENCODE and Umap100 exclude $<$4\% of genomic regions and are the most inclusive;
GIAB:strict and 1KG:strict exclude $>$20\% of genomic regions and are the smallest.
Our pm151 sets are something in between.
Most region sets include higher percentage of coding regions (CDS) than whole genome, which would be expected as CDS tends to be less repetitive.
Interestingly, Ilmn-easy misses $>$20\% of CDS albeit its high genome-wide coverage.
We suspect this is caused by the GC bias in short reads.
The Ilmn-easy regions were derived from the alignment of Illumina short reads.
Regions with read depth lower than 75\% of the genome average were excluded.
However, due to the sequencing bias against high-GC regions~\citep{Benjamini:2012aa} and the GC-rich nature of human CDS,
CDS would tend to have lower coverage and get excluded.
The 1KG easy regions were constructed in a similar way but with a read-depth threshold of 50\% instead of 75\%.
They are less affected by the coverage bias.
GIAB:strict excludes regions of extreme GC and is also biased against CDS.

To measure the coverage of SNP catalogs, we acquired SNPs of allele frequency of $\ge$10\% from gnomAD v4.1~\citep{Karczewski:2020aa}.
There are 5.85 million SNPs that have passed the gnomAD filters.
96.9\% of them are also called from 462 HPRC assemblies.
They are likely to be correct.
pm151 missed $\sim$10\% of these common SNPs (Fig.~\ref{fig:cov}a).
This suggests that taking full advantage of input data,
sophisticated SNP calling pipelines still have more power over easy regions.
On the other hand, we see a sharp difference in the fraction of gnomAD-filtered SNPs:
only 1.2\% of candidate SNPs inside pm151:strict are filtered by gnomAD,
but the percentage rises to 16.6\% for candidate SNPs outside pm151:strict.
gnomAD also thinks SNPs outside easy regions are harder to call.

We next compared the SNP and indel density using the GIAB Q100 callset.
This callset was generated from the near complete assembly of HG002~\citep{Rautiainen:2023ab}.
It contains variant calls across the whole genome.
Although calls outside the HG002-Q100 confident regions are less reliable,
most of these calls should also be outside easy regions and would not greatly affect the density analysis.
We note that while the SNP density is comparable across easy regions,
the indel density varies greatly (Fig.~\ref{fig:cov}b).
This is driven by the content of short tandem repeats (STRs) in easy regions as STRs harbor most of indels.
GIAB:strict explicitly filters out STRs and has the lowest indel density.
pm151:strict filters long LCRs and has reduced density as well.
The lower indel density in 1KG:strict might be related to its stringent mapping quality filtering
as BWA-MEM, the algorithm used for alignment, tends to give low mapping quality in long LCRs.

To measure the ease of calling, we compared the short-read variant calls by \citet{Baid2020.12.11.422022} to the GIAB Q100 callset
Importantly, although Q100 calls could be wrong outside the HG002-Q100 confident regions,
they are still more accurate than short-read calls
and even when Q100 calls are wrong, their consistency with short-read calls
still reflects the ``easiness'' of easy regions.

\begin{figure*}[tb]
\includegraphics[width=\textwidth]{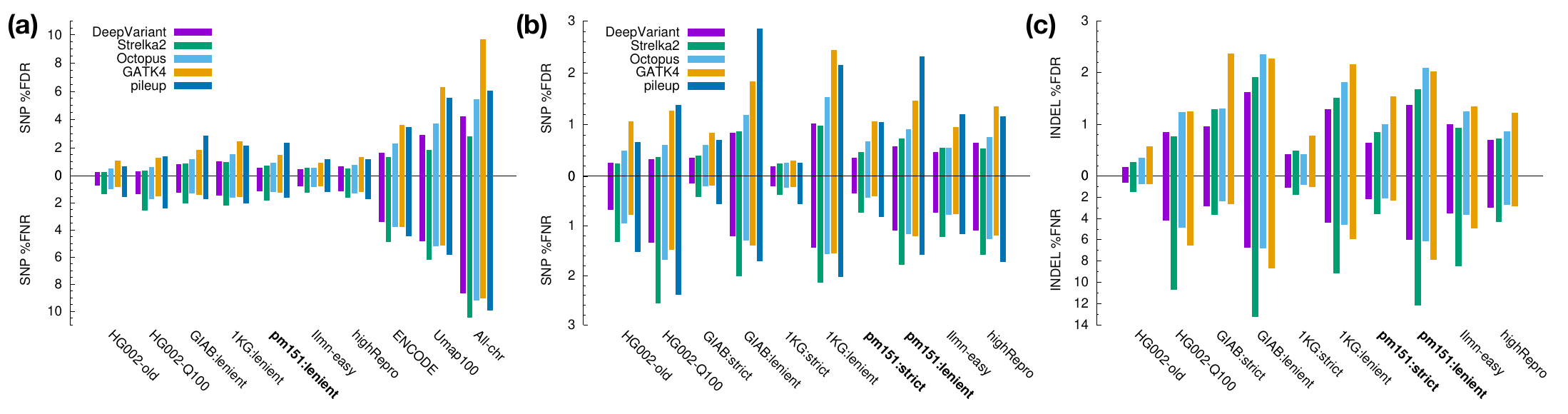}
\caption{Small variant calling accuracy in easy and confident regions.
{\bf (a)} SNP accuracy.
DeepVariant, Strelka2, Octopus and GATK short-read variant calls from 30X NovaSeq data were obtained from \citet{Baid2020.12.11.422022}.
Pipelup calls were made with ``{\tt minipileup -ys3 -a1 -p.25}''.
False discovery rate (FDR) and false negative rate (FNR) were calculated by RTG vcfeval~\citep{Cleary023754}
with genotype errors ignored.
``All-chr'' represents all chromosomal sequences in GRCh38 without easy regions.
{\bf (b)} SNP accuracy in high-quality easy regions.
{\bf (c)} Indel accuracy in high-quality easy regions.
Pileup indel calls are omitted due to their high error rates.
}\label{fig:ease}
\end{figure*}

We can see that the SNP calling error rates reach $\sim$10\% when no easy regions are applied (``All-chr'' in Fig.~\ref{fig:ease}a)
but are reduced by several folds to a couple of percent in GIAB, 1KG, pm151 and Ilmn-easy regions.
This sharp contrast manifests the necessity of easy regions.
Umap100 only considers exact 100-mer matches and overlooks the fact that high-identity inexact matches
are still challenging for aligners to distinguish.
It is inadequate as easy regions.
ENCODE blacklist is better than Umap100 but is still noticeably harder than other easy regions.

1KG:strict consists of the easiest regions (Fig.~\ref{fig:ease}b and~\ref{fig:ease}c).
It is almost a subset of pm151:lenient with 99.3\% included in the latter.
This overlap is higher than between 1KG:strict and Ilmn-easy (97.3\%) or GIAB:lenient (98.6\%).
Excluding long LCRs,
pm151:strict greatly reduces SNP and indel calling errors in comparison to pm151:lenient
and achieves a reasonable balance between coverage and easiness.

\section{Discussions}

The pm151 easy regions are used for filtering spurious variant calls in centromeres, long repeats,
or other genomic regions where short-read mapping is likely problematic.
These easy regions are not biased towards existing short-read data or aligners in use.
Given a prebuilt ropebwt3 index, they can be generated in two days for an arbitrary human assembly on a server with 64 CPU threads, with most time spent on ropebwt3 alignment.
The procedure can also be applied to a species with multiple well assembled genomes.

Overall, no easy region set is better than the others on all metrics.
Nevertheless, for the purpose of variant filtering,
we do not recommend GIAB:strict and Ilmn-easy as they miss $\sim$20\% of CDS and ClinVar pathogenic variants;
we discourage the use of ENCODE and Umap100 as they are too permissive
and are not effective for variant filtering.
1KG and pm151 are the better choices
though the 1KG set is only available to GRCh38.

We also caution that although easy regions provide a convenient way to identify high-quality variant calls,
they are conservative due to their generality.
They do not fully replace other variant filtering strategies.
Sophisticated variant calling pipelines optimized for input data and the best aligners and callers can achieve higher power and call more SNPs to high accuracy.

\section*{Acknowledgments}

We are grateful to Arang Rhie and Yoo-Jin Ha for pointing us to easy regions
derived by Illumina and the 1000 Genomics Project, respectively.
We would like to acknowledge the National Genome Research Institute (NHGRI) for
funding the following grants supporting the creation of the human pangenome
reference: U41HG010972, U01HG010971, U01HG013760, U01HG013755, U01HG013748,
U01HG013744, R01HG011274, and the Human Pangenome Reference Consortium
(BioProject ID: PRJNA730823).

\section*{Author contributions}

H.L. conceived the project, implemented the method, analyzed the data and drafted the manuscript.

\section*{Conflict of interest}

None declared.

\section*{Funding}

This work is supported by National Institute of Health grant R01HG010040, U01HG013748 and U41HG010972 (to H.L.).

\section*{Data availability}

pm151 easy regions can be downloaded from \url{https://doi.org/10.5281/zenodo.14903542}. Scripts to generate easy regions are available at \url{https://github.com/lh3/panmask}.

\bibliographystyle{apalike}
{\sffamily\small
\bibliography{panmask}}

\end{document}